\documentclass[11pt, notoc,letter]{JHEP3}
\usepackage{graphicx}
\usepackage{amssymb}
\usepackage{amsmath}
\usepackage{latexsym}

\newcommand{\be}{\begin{equation}}
\newcommand{\beq}[1]{\begin{equation}\label{#1}}
\newcommand{\eq}[1]{Eq.~(\ref{#1})}
\newcommand{\ee}{\end{equation}}
\newcommand{\eeq}{\end{equation}}
\newcommand{\bea}{\begin{eqnarray}}
\newcommand{\eea}{\end{eqnarray}}
\newcommand{\comment}[1]{}
\newcommand{\TeV}{~\mathrm{TeV}}

\newcommand{\Pl}{\mathrm{Pl}}
\newcommand{\lsim}{\!\mathrel{\hbox{\rlap{\lower.55ex \hbox{$\sim$}} \kern-.34em \raise.4ex \hbox{$<$}}}}
\newcommand{\gsim}{\!\mathrel{\hbox{\rlap{\lower.55ex \hbox{$\sim$}} \kern-.34em \raise.4ex \hbox{$>$}}}}

\DeclareMathOperator{\Rej}{Re}

\newcommand\openone{\leavevmode\hbox{\small1\normalsize\kern-.33em1}}

\title{Prospects for Mirage Mediation}
\author{Aaron Pierce and Jesse Thaler \\ Jefferson Physical Laboratory, Harvard University, Cambridge, MA 02138 \\ E-mail:  \email{apierce@physics.harvard.edu, jthaler@jthaler.net}}
\preprint{HUTP-06/A0013}

\abstract{Mirage mediation reduces the fine-tuning in the minimal supersymmetric standard model by dynamically
arranging a cancellation between anomaly-mediated and modulus-mediated
supersymmetry breaking.  We explore the conditions under which a mirage ``messenger scale'' is generated near the weak scale and the little hierarchy problem is solved.  We do this by explicitly including the dynamics of the SUSY-breaking sector needed to cancel the cosmological constant.  The most plausible scenario for generating a low mirage scale does not readily admit an extra-dimensional interpretation.  We also review the possibilities for solving 
the $\mu/B\mu$ problem in such theories, a potential hidden source of 
fine-tuning.}

\begin{document}

\section{Motivation}
For twenty years, supersymmetry (SUSY) has reigned as the most compelling solution to the hierarchy problem.  Recently, confidence that the minimal supersymmetric standard model (MSSM) is realized in nature has been eroded by the so-called ``little hierarchy problem''.   In the MSSM, the little hierarchy appears as a tension between the two different roles played by the stop squark:  on the one hand, the stop should be light  to cut off the quadratic divergence in the Higgs boson (mass)$^{2}$ and thus minimize fine-tuning; on the other hand, the stop should be heavy enough to generate a physical mass for the Higgs boson above the LEP II bounds.  This tension is exacerbated by renormalization group (RG) flow which tends to drive colored particles heavier than electroweak particles.  The LEP II bound on charginos, along with the assumption of universal gaugino masses, generically ensures that the stop is quite heavy, requiring fine-tuning at the percent level to achieve successful electroweak symmetry breaking.

One potential way to reduce the level of fine-tuning is to have a compressed SUSY spectrum at low scales.  Any mechanism yielding such a spectrum while simultaneously evading the bound on the Higgs boson mass merits study.  Mirage mediation \cite{Choi:2005hd,Kitano:2005wc}, based on the investigations of \cite{Choi:2004sx,Choi:2005ge}, claims to be such a mechanism.  In mirage mediation, a cancellation between modulus mediation \cite{Kaplunovsky:1993rd,Brignole:1993dj} and anomaly mediation \cite{Randall:1998uk,Giudice:1998xp} creates a distinctive low energy SUSY spectrum that appears to unify at a mirage ``messenger scale'' $M_{\rm mess} \ll M_{\rm GUT}$.  Because this cancellation occurs at every RG scale, there is no hidden fine-tuning between the modulus- and anomaly-mediated contributions \cite{Kitano:2006gv}.  In addition, mirage mediation generically yields large trilinear $A$ terms which generate a large enough physical Higgs mass.  Since the leading correction to the Higgs (mass)$^{2}$ goes like
\begin{equation}
\Delta m_{H_{u}}^{2} = -\frac{3 y_{t}^{2}}{4 \pi^{2}} m_{\tilde{t}}^{2} \; {\rm ln} \left( \frac{M_{\rm mess}}{m_{\tilde{t}}}\right), 
\end{equation}
the level of fine-tuning can be reduced if $M_{\rm mess}$ is close to the weak scale.   

The value of $M_{\rm mess}$ is determined by the relative contributions to the visible sector soft masses of modulus and anomaly mediation.  However, this ratio can only be evaluated after SUSY has been broken and the cosmological constant (CC) has been tuned to zero.  Mirage mediation occurs in theories where a modulus $T$ is stabilized in a supersymmetric AdS vacuum uplifted by a independent source of SUSY breaking.   In the simplest theory where a mirage scale can be realized, the KKLT construction \cite{Kachru:2003aw}, the AdS vacuum is lifted by the presence of a SUSY-breaking $\overline{D3}$ brane.   In this case, one finds $M_{\rm mess} \sim \sqrt{m_{3/2} M_\Pl}$, which would not solve the little hierarchy problem.  To achieve a lower mirage messenger scale, the authors of \cite{Choi:2005hd,Kitano:2005wc} break SUSY explicitly through an ad-hoc uplift potential proportional to $(T+T^\dagger)^{n_{\rm lift}}$, where $n_{\rm lift}$ is assumed to take on discrete values.  The mirage messenger scale is then found to be
\be
M_{\rm mess} \sim \frac{M_{\rm GUT}}{\left(M_\Pl / m_{3/2}\right)^{\alpha/2}},
\ee
where $\alpha$ is a function of $n_{\rm lift}$.  Written this way, it appears that $M_{\rm mess} \sim$ TeV can be achieved simply by assuming $\alpha = 2$.  

However, because the modulus is stabilized without breaking supersymmetry, the uplift sector is the dominant source of SUSY breaking in mirage mediation.  As discussed in \cite{Lebedev:2006qq}, the goldstino has a large overlap with the uplift sector, so the ad-hoc uplift potential neglects the dynamics of light fields in the complete theory.  In this paper, we assess the prospects for mirage mediation to solve the little hierarchy problem by attempting to construct an explicit model where $M_{\rm mess} \sim$ TeV.  We find that the most plausible scenario for achieving this low $M_{\rm mess}$ is for SUSY breaking to be triggered by a chiral superfield $X$ with effective modular weight $r_X$, yielding
\be
\alpha = \frac{2}{2 + r_X}.
\ee
If $T$ is a volume modulus, $r_X$ is expected to range between $0$ and $1$, and for brane-localized SUSY breaking as in KKLT, $r_X = 0$ and thus $\alpha=1$.  To get $\alpha = 2$ requires $r_X = -1$, so while there is indeed a 4D supergravity Lagrangian that yields a low mirage messenger scale, this theory has some peculiar properties.  As $T$ grows large gravity gets weaker, but negative $r_X$ implies that the uplift sector must get strongly coupled in the decompactification limit.   We conclude that a crucial model-building challenge to achieving mirage mediation with low fine-tuning is to find a justification for negative effective modular weights in the uplift sector.

Furthermore, before any supersymmetric model can declare victory over the fine-tuning problem, it must demonstrate a plausible mechanism for generating the right $\mu/B\mu$ ratio.  This issue is of special concern for mirage mediation because $m_{3/2} \gg M_{Z}$.  As in any theory where anomaly mediation is active, one generically expects $B \sim m_{3/2}$, resulting in a fine-tuning for the weak scale.  Interestingly, we find that the only solution in the literature that seems to work without additional fine-tuning explicitly relies on the existence of a low mirage messenger scale \cite{Choi:2005hd}.


In the next section, we review the mechanism of mirage mediation in the language of \cite{Choi:2005hd,Kitano:2005wc} where $M_{\rm mess}$ is determined by an ad-hoc uplift 
potential.  In Section \ref{sec:uplift}, we assess the plausibility of $M_{\rm mess} \sim \mbox{TeV}$ by explicitly include the source of SUSY breaking that generates the uplift potential.  The issue of $\mu/B\mu$ is examined in Section \ref{sec:muBmu}.   We conclude with some comments about mirage mediation and cosmology.




\section{Review of Mirage Mediation}
\label{sec:review}
The discussion in this section closely follows \cite{Kitano:2005wc,Choi:2004sx,Choi:2005ge,Endo:2005uy}.  In the conformal compensator language, the leading two-derivative supergravity Lagrangian can be 
written as \cite{conformal}
\be
\label{eq:sugralang}
\mathcal{L} = \int \! d^4 \theta \,  \phi^\dagger \phi \, \Omega + \int \! d^2  \theta\,  \phi^3 \, W + \mbox{h.c.} + \mbox{gravity and gravitino terms}
\ee
with $\phi= 1 + F_{\phi} \theta^{2}$.  The starting point for mirage mediation is a modulus $T$ with a 
KKLT-like K\"ahler potential and superpotential:
\be
\label{Eqn: KKLTW}
\Omega \sim (T+T^\dagger)^{n_0}, \qquad W = ae^{-bT} + c,
\ee
where $b$ is a real coefficient, and one expects
\be
a \sim M_\Pl^3, \qquad c \sim m_{3/2} M_\Pl^2, \qquad m_{3/2} \sim 10-100 \TeV.
\ee
There are two contributions to the SUSY-breaking masses in the visible sector.  First, there is the contribution from anomaly mediation, which is suppressed by a loop factor from the scale $F_{\phi}$.  The second contribution is proportional to
\be
M_0 = \frac{F_T}{T+T^\dagger},
\ee
which parameterizes the supersymmetry breaking mediated by the modulus $T$.  

A mirage messenger scale is generated if the couplings of the visible sector fields to the modulus $T$ obey certain properties. The modulus field must couple linearly to the gauge kinetic terms:
\be
 \int d^{2} \theta \, \frac{T}{4}  {\mathcal W}^{\alpha}_{a}{\mathcal W}_{\alpha}^{a}.
\ee
The couplings of $T$ to the chiral superfields of the MSSM are also constrained.  A generic visible sector field $Q_{i}$ appears in the K\"ahler potential as
\be
\Omega \ni Q^{\dagger}_{i} Q_{i} (T+ T^{\dagger})^{n_0 r_{i}},
\ee
with $r_{i}$ giving the effective modular weight of the field $Q_{i}$. Then, if the superpotential contains
\be
W \ni \lambda_{ijk} Q_{i} Q_{j} Q_{k}
\ee  
with an appreciable Yukawa coupling, the existence of a well-defined mirage scale is guaranteed if
\be
r_{i} + r_{j} + r_{k}= 1.
\ee
Furthermore, the hypercharges of the chiral superfields must obey the relation
\be
\sum_i \; r_{i} Y_{i}  = 0.
\ee
As emphasized in \cite{Kitano:2005wc}, these requirements on the modular weights can be realized by choosing $r_{M}=1/2$ for matter fields and $r_{H} =0$ for Higgs fields. This choice also avoids dangerous flavor changing neutral currents, and a geometric picture for obtaining these weights from $D7$ and $D3$ branes was presented in \cite{Kitano:2005wc}. 

If these conditions on the couplings of $T$ to the visible sector are met, then the modulus- and anomaly-mediated contributions to the visible sector soft masses exhibit a nice cancellation.   For example, including the effect of one-loop running, the gaugino masses at the RG scale $\mu_R$ are given by
\bea
m_a(\mu_R) &=& M_0 \left(1 - \frac{b_a g^2_a (\mu_R )}{8 \pi^2} \log \frac{M_{GUT}}{\mu_R} \right) + \frac{b_a g^2_a (\mu_R )}{16 \pi^2} F_\phi,\\
& = & M_0 \left(1 - \frac{b_a g^2_a (\mu_R )}{8 \pi^2} \log \frac{M_{\rm mess}}{\mu_R} \right),
\eea
where the mirage messenger scale is given by 
\be
M_{\rm mess} \equiv M_{\rm GUT} \, e^{- F_\phi / 2 M_0}.
\ee
Of course, the mere existence of a mirage scale is insufficient to solve the little hierarchy problem; the mirage scale should be also be near the weak scale.  The key realization is that as long as $F_\phi / M_0$ is sufficiently large, then $M_{\rm mess}$ can be $\mathcal{O}(\mbox{TeV})$.   A priori, we have no reason to expect $F_\phi / M_0$ to have any particular value, but given the superpotential in \eq{Eqn: KKLTW}, one expects
\be
\label{eq:naturalbtsize}
bT \sim \log \frac{a}{c} \sim \log \frac{M_\Pl}{m_{3/2}}.
\ee
Therefore, if $F_\phi / M_0$ can be calculated as a power series in this quantity
\be
\label{eq:ratioinepsilon}
\frac{F_\phi}{M_0} = \frac{\alpha}{\epsilon} + \mathcal{O}(\epsilon^0), \qquad \epsilon \equiv \frac{1}{\langle b\Rej{T} \rangle},
\ee
then we can have a parametrically small $M_{\rm mess}$ if $\alpha$ is non-zero. That is to say, the hierarchy needed between $F_{\phi}$ and $M_{0}$ can be related to the stabilization of $b T$ at large values.  In particular, if \eq{eq:naturalbtsize} holds,\footnote{For any non-zero value of $\alpha$, one can of course fine-tune the ratio $F_\phi/M_0$ by adjusting the values of $a$ and $c$.  However, absent a good reason to expect $a/c$ to be substantially different from $M_\Pl/m_{3/2}$, we ignore this mechanism for adjusting $M_{\rm mess}$.  Similarly, if $M_\Pl$ is determined by other moduli in addition to $T$, then the equations of motion for $T$ will no longer set $bT \sim \log M_\Pl / m_{3/2}$ (see \eq{eq:appendixbtvalue}) and we lose the predictive power of mirage mediation.} 
\be
M_{\rm mess} \sim \frac{M_{\rm GUT}}{\left(e^{1/\epsilon} \right)^{\alpha/2}} \sim \frac{M_{\rm GUT}}{\left(M_\Pl / m_{3/2}\right)^{\alpha/2}}.
\ee
Once a mechanism is specified to obtain $\alpha \simeq 2$, then $M_{\rm mess} \sim \mbox{TeV}$ as desired for the little hierarchy problem.

The value of $\alpha$ depends on the way in which the AdS vacuum generated by \eq{Eqn: KKLTW} is lifted.   In \cite{Choi:2005hd,Kitano:2005wc}, the uplift potential is included by adding an explicit SUSY breaking term to the K\"ahler potential
\be
\Omega_{\rm lift} = \theta^{2} \bar{\theta}^{2} V_{\rm lift}, \qquad V_{\rm lift} = d (T+T^\dagger)^{n_{\rm lift}}.
\ee
Though the origin of $V_{\rm lift}$ is not specified by \cite{Choi:2005hd,Kitano:2005wc}, they note that if $n_{\rm lift} = n_0$, then $\alpha = 2$ after tuning the CC to zero.

Already at this level, this uplift potential looks strange.  Because $T$ is a modulus, \eq{Eqn: KKLTW} suggests that $n_0$ should be positive.  On the other hand, $V_{\rm lift}$ characterizes the size of SUSY breaking, and one does not normally expect SUSY breaking to grow with volume, as would be the case for $n_{\rm lift} = n_0$.  For brane-localized SUSY breaking, $V_{\rm lift}$ is expected to be independent of $T$.  Therefore, a complete theory that yields $M_{\rm mess} \sim \mbox{TeV}$ will probably not have a nice extra-dimensional interpretation.  

\section{Assessing the Naturalness of the Uplift Potential}
\label{sec:uplift}

Given how important the form of the uplift potential is for setting $M_{\rm mess}$, we want to understand how to generate such a potential from a complete theory.  In particular, the $T$ sector does not break SUSY on its own, so whatever field(s) contribute to the SUSY-breaking uplift potential will have a large overlap with the goldstino.  As recently discussed in \cite{Lebedev:2006qq}, instead of simply inserting a SUSY-breaking uplift spurion, one should keep track of the light fields in the uplift sector.



The best prospect for finding a field-theoretic realization of the desired uplift potential is a 4D, $N=1$ theory.  Additional constraints arising from $N=2$ SUSY in a bulk 5D theory could only make constructing the desired uplift potential more difficult. The case of a single chiral field $X$ should be general if we impose that there be no fine-tunings to realize the uplift potential; additional fields would necessitate tunings between different SUSY breaking contributions.  We allow arbitrary mixing between $X$ and $T$ in the K\"ahler potential as long as the shift symmetry on $T$ is preserved, and we allow arbitrary mixing in the superpotential as long as $T$ appears only once in the combination $e^{-bT}$.  Direct mediation of SUSY breaking to MSSM multiplets $Q_i$ from K\"ahler potential terms like $X^\dagger X Q_i^\dagger Q_i / M_\Pl^2$ would spoil the predictions of mirage mediation, so these terms must  be suppressed via some form of sequestering.  With these assumptions, our effective 4D theory is given by
\be
\label{eq:xsetup}
\Omega = \Omega(X, X^\dagger, T+T^\dagger), \qquad W = a(X) e^{-bT} + c(X),
\ee
where $\Omega$, $a$, and $c$ are arbitrary functions and $b$ is a real parameter.

In Appendix \ref{app:alpha}, we calculate $\alpha$ from \eq{eq:ratioinepsilon} using the specific form of \eq{eq:xsetup}.  To leading order in $\epsilon$, tuning the CC to zero is equivalent to sending
\be
\label{eq:cctune}
\frac{F_\phi}{h F_X} \rightarrow -1, \qquad h = \frac{c_X}{3c},
\ee
where subscripts on functions indicate partial derivatives and all functions are evaluated at the minimum of the potential.  For reference:
\be
\label{eq:fphifxratio}
\frac{F_\phi}{F_X} = \frac{h^* \Omega_X - \Omega_{X^\dagger X}}{\Omega_{X^\dagger} - h^* \Omega} + \mathcal{O}(\epsilon).
\ee
Once the CC tuning is applied, we find
\be
\label{eq:alphabetagamma}
\alpha = \frac{2}{2-\beta - \gamma}, \qquad \beta = \frac{1}{h} \Bigl( \frac{h^* \Omega_{XT} - \Omega_{X^\dagger X T}}{\Omega_{X^\dagger T}-h^*\Omega_T}\Bigr), \qquad \gamma = \frac{1}{h}\frac{a_X}{a}.
\ee
We see immediately that the only way to get make $\alpha \neq 1$ is to have terms in $\Omega$ or $W$ that mix $X$ and $T$.  In the original KKLT scenario \cite{Kachru:2003aw}, the role of $X$ is played by a $\overline{D3}$ brane which does not communicate with the modulus field $(\beta=\gamma=0)$, generically leading to $\alpha = 1$.

The philosophy of mirage mediation is that $\alpha$ should be determined by discrete choices in the Lagrangian.  In this way, determination of $\alpha$ does not contribute to the fine-tuning present in the theory.  Since the expression for $\alpha$ in \eq{eq:alphabetagamma} depends on various combinations of functions and their derivatives, avoiding explicit tunings between terms requires focusing on limits where a given term dominates.   We will start by checking whether $\gamma$ can naturally take on discrete values, then we will consider different options for adjusting $\beta$. Because $\Omega_{XT}$ appears in both the numerator and denominator of $\beta$, there are three cases that avoid an explicit tuning:  (\emph{i}) the tuning of the CC naturally relates the numerator and denominator of $\beta$, (\emph{ii}) $\Omega_{XT}$ dominates, or (\emph{iii}) $\Omega_{XT}$ is very small.  At the end of the day, this third method will yield a viable model with $\alpha = 2$, but the form of the K\"ahler potential will not have an obvious extra-dimensional interpretation.  

First, from \eq{eq:alphabetagamma} it is possible for $\gamma$ to take on discrete values if 
\be
\label{Eq:peculiarW}
a(X) = A^{1-n_a} c(X)^{n_a}, 
\ee
where $A$ has mass dimension $3$.  This yields $\gamma = 3 n_a$.  Not only is it unclear how to obtain such a superpotential from a top-down theory, but there is a second problem.  The goal of mirage mediation is to generate $F_{\phi}/M_{0} \sim 2 \log M_{\Pl} /m_{3/2}$.  While this special form of the superpotential assures discrete choices for $\alpha$, it no longer guarantees that $\langle b T \rangle$ will be quantitized in units of $\log M_{\Pl}/m_{3/2}$,  so there is no sense in which  $F_{\phi}/M_{0}$ is a discrete choice.  In particular,
\be
\langle b T \rangle \sim \log \frac{a}{c} \sim (1-n_a) \log \frac{A}{c}, 
\ee
and the stabilization of $b T$ depends on the choice of $A$.  If one could motivate a particular value for $A$, it might be interesting to see whether the peculiar superpotential in \eq{Eq:peculiarW} might lead to weak scale mirage mediation.  But for the remainder of the section, we will take unceratinty in $A$ as an indicator of fine-tuning and assume $\gamma = 0$.

How can we arrange $\beta$ to take on discrete values?  Note that $\beta h$ in \eq{eq:alphabetagamma} looks very similar to $F_\phi / F_X$ in \eq{eq:fphifxratio} with additional $T$ derivatives.  Because the tuning of the CC in \eq{eq:cctune} already relates $h$ to $F_\phi / F_X$, it seems plausible that this tuning might be sufficient to insure special values for $\beta$.   
However, for generic values of $\Omega_{XT}$, the only time we can make use of the CC tuning is when the $T$ derivatives in \eq{eq:alphabetagamma} act the same way on each term in $\beta$.  In particular,
\be
\frac{\Omega_{T}}{\Omega} = \frac{\Omega_{XT}}{\Omega_{X}} = \frac{\Omega_{X^\dagger X T}}{\Omega_{X^\dagger X}} = f(T+T^\dagger), \qquad \beta = \frac{F_\phi}{h F_X} = -1, \qquad \alpha = \frac{2}{3}.
\ee
This value of $\alpha$ does not substantially improve the little hierarchy problem.

A more promising method to get $\beta \simeq 1$  ($\alpha \simeq 2$) is to let $h^* \Omega_{XT}$ be real  and large.  In particular, $\alpha \rightarrow 2$ in the limit
\be
\label{eq:omegaxtcondition}
h^* \Omega_{XT} \gg \Omega_{X^\dagger X T}, \qquad h^* \Omega_{XT} \gg |h|^2 \Omega_T.
\ee
However, the first of these conditions seems problematic.  If $X$ starts off with a kinetic term $X^\dagger X (T+T^\dagger)^{n_X}$, we expect $\Omega_{XT} \sim \langle X^\dagger \rangle \Omega_{X^\dagger X T}$.  Once we tune the CC, the first condition in \eq{eq:omegaxtcondition} tells us
\be
\langle X^\dagger \rangle \gg \frac{F_X}{F_\phi} \sim M_\Pl,
\ee
and one worries about the reliability of effective field theory in the presence of Planckian vevs.  

Now, we {\it can} achieve $\alpha = 2$ consistent with the philosophy of mirage mediation if we take $\Omega_{XT} \rightarrow 0$.   In this case, the leading K\"ahler potential should have the form
\be
\label{eq:noomegaxt}
\Omega = -d_1 (T+T^\dagger)^{n_0} + d_2 X^\dagger X (T+T^\dagger)^{r_X n_0} + \cdots,
\ee
where $d_i$ are positive constants of the appropriate dimensionality.  In order for $\Omega_{XT} \sim 0$, there must be some reason for $\langle X \rangle = 0$, and the only symmetry that allows $F_X \not= 0$ with $\langle X \rangle = 0$ is an $R$-symmetry where $X$ has $R$-charge $2$.  As shown in Appendix \ref{app:alpha}, the $R$-breaking induced by $c$ generically shifts $\langle X \rangle$ by a controllable amount that does not drastically change $\alpha$.  Aside from those small corrections, we find
\be
\beta = -r_X, \qquad \alpha = \frac{2}{2 + r_X},
\ee
so it is indeed possible to get $\alpha = 2$ if $r_X = -1$.  

Unfortunately, we know of no geometric interpretation to justify negative $r_X$ values.  Negative $r_X$ implies that as $T \rightarrow \infty$, $X$ gets more and more strongly coupled, precisely the opposite behavior of ordinary fields propagating in extra dimensions.  Note that $X^\dagger X / (T+T^\dagger)$ must generate the leading kinetic term for $X$; any bare $X^\dagger X$ piece will drag $\alpha$ smaller than 2.   Therefore, a crucial model building challenge to achieving mirage mediation with low fine-tuning is finding a justification for $X$ to have effective modular weight $r_X = -1$.  Also, the uplift sector must be sequestered from the visible sector in order to suppress the contribution of higher dimension operators to the MSSM soft masses, but the standard method of brane-localization is not compatible with $r_X = -1$.  However, apart from questions of plausibility, the model in \eq{eq:noomegaxt} has no tachyons or ghosts and all fields can be stabilized.


\section{The $\mu$/$B\mu$ Problem in Mirage Mediation}
\label{sec:muBmu}

The origin of $\mu$ term presents a puzzle: why should this supersymmetric term be of order the weak scale? As with 
all theories with large gravitino mass, this question is sharpened in the case of mirage mediation.  
The simplest supergravity solution, the Giudice--Masiero (GM) mechanism \cite{GiudiceMasiero}, is unavailable. 
Because the gravitino mass is a loop-factor above the weak scale, GM predicts a too 
large $\mu \sim m_{3/2} \gg M_{Z}$.  One can alleviate this issue by requiring the coefficient 
of the GM operator to be small, but another problem remains. The contribution to 
$B\mu$ arising from the conformal compensator is generically of 
order $F_{\phi} \mu$, yielding $B \sim m_{3/2}$.   Since we require $|m_{H_{u}}^{2}|,|m_{H_{d}}^{2}| \sim M_{Z}^{2}$ to keep fine-tuning small, this would prevent proper electroweak symmetry breaking.  In theories like  mirage mediation where anomaly mediation is active, the challenge is to ensure the generated $B$-term is sufficiently small.  In this section, we review three possibilities for generating the $\mu$ and $B\mu$ terms, and assess the prospects for each mechanism. 

One possibility, originally discussed 
in \cite{Randall:1998uk} and adopted 
for the case of mirage mediation in \cite{Kitano:2005wc}, is to forbid the 
leading GM operator, and instead rely on the operator:
\begin{equation}
\label{Eqn: Solvemu}
\mathcal{L}_{\mu} = \int d^{4} \theta \; \phi^{\dagger} \phi \, \left( y (\Sigma +\Sigma^{\dagger}) H_{u} H_{d} + \rm{h.c.} \right),
\end{equation}
where $\Sigma$ is normalized to be dimensionless and $y$ is a dimensionless coupling.  This operator generates 
\begin{eqnarray}
\label{Eqn: muBmu}
\mu &=& y \left(F_{\Sigma}^{\dagger} +  F_{\phi}^{\dagger} \langle \Sigma +\Sigma^{\dagger} \rangle\right), \\
B \mu &=& - y \left( |F_{\phi}|^2 \langle\Sigma+ \Sigma^{\dagger}\rangle + (F_{\Sigma} F_{\phi}^{*} -{\rm h.c}) \right) + (\gamma_{H_{u}} + \gamma_{H_{d}}) F_{\phi} \mu.
\end{eqnarray}
The last term in the expression for $B\mu$ is the usual anomaly--induced contribution.  It comes from the wave-function renormalization of the $H_{u}$ and $H_{d}$ fields, analytically continued into superspace.   After RG evolution down to the weak scale, this final term can give $\mu$ and $B\mu$ in the proper ratio \cite{Kitano:2005wc}.  

Thus, to utilize the above mechanism we need to be able to neglect the first two contributions to $B\mu$.  What is the generic expectation for the size of these terms?  First, assume that $\langle \Sigma + \Sigma^{\dagger} \rangle$ is negligible.  Then, working in the basis where $F_{\phi}$ is real, we can insure that $B\mu$ is dominated by the anomaly-mediated piece if $F_{\Sigma}$ is real.  This choice has the added benefit of solving the SUSY CP problem \cite{Endo:2005uy}.  The $\mu$ term can be taken real, and there is no relative phase between $\mu$ and $B\mu$.  Furthermore, in mirage mediation the gaugino masses and the $A$-terms are proportional to $F_\phi$, so when $F_\Sigma$ is real, there are no relative phases between these soft terms and $\mu$.  

Unfortunately, it is difficult to make $\langle \Sigma + \Sigma^{\dagger} \rangle$ vanish while simultaneously keeping $F_{\Sigma}$ real. The term in \eq{Eqn: Solvemu} has a special form, dependent on the combination $\Sigma + \Sigma^{\dagger}$.  An operator of this form most naturally would appear in theories where $\Sigma$ is modulus and therefore obeys a shift symmetry.   References \cite{Choi:1993yd,Endo:2005uy} studied the conditions under which a real $F_{\Sigma}$ could be ensured for a modulus field.   Taking into account the equation of motion for Im($\Sigma$), the condition for a real $F_{\Sigma}$ is 
\begin{equation}
\frac{W_{\Sigma \Sigma}}{W_{\Sigma}} \in \mbox{Reals}.
\end{equation} 
This requires a superpotential of the form \cite{Choi:1993yd,Endo:2005uy}
\begin{equation}
W_{\Sigma} =  a_{\Sigma} e^{-b_{\Sigma} \Sigma} + c_{\Sigma},
\end{equation}
with $b_{\Sigma}$ real.  However, this superpotential does not allow stabilization of the $\Sigma$ field at small vev for generic coefficients.  In fact, from the discussions in Section \ref{sec:review}, we know that unless $c_{\Sigma}$ and $a_{\Sigma}$ are fine-tuned, this superpotential leads to
\begin{equation}
\frac{F_{\Sigma}}{\langle \Sigma+\Sigma^{\dagger} \rangle} \ll F_{\phi}.
\end{equation}   
But from  \eq{Eqn: Solvemu}, a non-zero vev for $\langle \Sigma + \Sigma^{\dagger} \rangle=0$ is equivalent to introducing a GM term.  So, once a non-zero vev is induced, we can recover the right $\mu$/$B\mu$ ratio, but only at the expense of fine-tuning against a bare GM term.  Since the hope was that mirage mediation might give rise to a minimally tuned MSSM, this seems to present a strong challenge to this mechanism for $\mu/B \mu$.  

One might wonder whether imposing this shift symmetry on $\Sigma$ is too restrictive.  
However, the difficulty in keeping $\langle \Sigma + \Sigma^{\dagger} \rangle =0$ while inducing a non-zero $F_{\Sigma}$ seems to be more generic.  Given the goal of achieving a minimally fine-tuned MSSM, it is natural to impose a symmetry to keep  $\langle \Sigma + \Sigma^{\dagger} \rangle = 0$.  However, the only symmetry that make $\Sigma  = 0$ special while still allowing $F_\Sigma  \not= 0$ is an $R$-symmetry where $\Sigma$ has $R$-charge $2$, but this $R$-symmetry is inconsistent with the K\"ahler potential in \eq{Eqn: Solvemu}.\footnote{One might be able to construct a theory with a spontaneously broken $\Sigma \rightarrow -\Sigma$ symmetry, but we do not know of an explicit model.} We conclude that the mechanism of \eq{Eqn: Solvemu} does not appear to solve the $\mu$ problem without additional fine-tuning.

A second potential solution to the $\mu$ problem is the Next to Minimal Supersymmetric Standard Model (NMSSM) \cite{NMSSM}. In the context of mirage mediation, this solution was advocated in \cite{Choi:2005uz}.   Unfortunately. the NMSSM superpotential
\begin{equation}
W = \lambda N H_{u} H_{d} + \kappa N^{3}
\end{equation}
is inconsistent with the stringent requirements necessary to keep the RG flow on the mirage-mediated trajectory.   Recall, if the superpotential contains a term $\phi_{i} \phi_{j} \phi_{k}$, the corresponding effective modular weights must satisfy 
\begin{equation}
\label{Eqn:SumCondition}
r_{i} + r_{j} +r_{k} =1.  
\end{equation}
But at the scale $M_{\rm mess}$, the soft scalar masses in mirage mediation are given by \cite{Choi:2005hd,Kitano:2005wc}
\be
m_i^2(M_{\rm mess}) = M_0^2 r_i.
\ee
Therefore, non-zero vacuum expectation values for $N$, $H_{u}$, and $H_{d}$ favor $r_{N} = r_{H_{u}} =r_{H_{d}}=0$, otherwise the fields have large positive $($mass$)^{2}$ at the weak scale.\footnote{We could try giving fields negative effective modular weights so that they have large negative $($mass$)^{2}$ at the weak scale, but this pushes us further off the mirage-mediated trajectory. Plus, as in Section \ref{sec:uplift}, we do not know of an extra-dimensional interpretation for negative effective modular weights.}  These two requirements are inconsistent, so any NMSSM solution to the $\mu$ problem will result in a deflected-mirage mediation spectrum.  Depending on the size of $\lambda$, this deflection could be significant.  

That said, even if we set $r_{N} = r_{H_{u}} =r_{H_{d}}=0$ and ignore the issue of deflection, it is not obvious that mirage mediation will yield the right soft masses for the NMSSM.  The calculable contribution to the soft masses of the Higgs bosons come from anomaly mediation.  AMSB suffers from the usual difficulty of theories with great predictive power on the soft masses---it is not automatic to get proper electroweak symmetry breaking while obtaining a large enough $\mu$ term.  In particular, in the minimal implementation of the NMSSM, the anomaly-mediated contribution to the $N$ soft mass is generically positive \cite{KKM}.  Therefore, in the absence of negative contributions to $m_{N}^{2}$ at the GUT scale, it might be necessary to introduce additional vector-like pairs coupled to $N$, perhaps along the lines of \cite{LutyChacko,Ami}.  If a convincing mechanism for $\alpha=2$ can be found, the phenomenology and model building possibilities for mirage mediation in the NMSSM might be worth pursuing further.


Finally, there is a promising mechanism to solve the $\mu/B\mu$ problem that is quite special to mirage mediation \cite{Choi:2005hd}.   It relies on the $\mu$ term being generated by the same underlying non-perturbative dynamics that lead to the superpotential of \eq{Eqn: KKLTW}.  If the $\mu$ term is generated by the superpotential:
\begin{equation}
\label{Eqn:SolveMuNP}
W_{\mu} = A e^{-b T} H_{u} H_{d},
\end{equation}
with $b$ identical to the $b$ of \eq{Eqn: KKLTW},
then the correct ratio of $\mu$ and $B \mu$ can be achieved if $\alpha = 2$.  This superpotential gives rise to
\begin{eqnarray}
\mu &=&  \langle A e^{-b T} \rangle\\
B \mu &=& -b F_{T} \langle A e^{-b T} \rangle + F_{\phi} \langle A e^{-b T} \rangle + (\gamma_{H_{u}}+\gamma_{H_{d}}) F_{\phi} \mu \\
      &=& \mu \left[  F_{\phi} -2 M_0 \langle b \Rej T \rangle \right] + (\gamma_{H_{u}}+\gamma_{H_{d}}) F_{\phi} \mu.
\end{eqnarray}
If the terms in brackets cancel, we are left with just the sub-dominant anomaly-mediated 
contribution, yielding the correct $\mu/B\mu$ ratio.  Note that
\be
F_{\phi} - 2 M_0 \langle b \Rej T \rangle = M_0 (\alpha -2 + \mathcal{O}(\epsilon)), 
\ee
so this cancellation only occurs when $\alpha=2$, and therefore relies on the special relationship between $M_0$ and $F_{\phi}$ when there is a low mirage scale.  

Interestingly, once a workable mechanism to achieve $\alpha=2$ is found, then it is feasible to generate a potential of the form $W_{\mu}$.  $T$ might be a modulus responsible for setting a high energy gauge coupling for a supersymmetric gauge theory.   For example, in pure SUSY Yang-Mills, an operator of the form $W_{\mu}$ could arise from a coupling in the high energy theory of the form
\be
 \int d^{2} \theta \, \frac{y}{M^2} {\mathcal W}^{\alpha} {\mathcal W}_{\alpha} H_{u} H_{d}.
\ee
There is gaugino condensation in the low energy theory, and the operator ${\mathcal W}^{\alpha}{\mathcal W}_{\alpha}$ may be replaced by $\Lambda^{3} e^{-b T}$.    Not only will gaugino condensation contribute to the above operator,  it will also give rise to the  $a e^{-b T}$ in \eq{Eqn: KKLTW} through the usual gauge kinetic term.  This determines the necessary size of $y/M^{2}$.  In particular, as discussed in Section~\ref{sec:review}, $a e^{-bT} \sim \Lambda^{3} e^{-bT}$ is expected to be of order $m_{3/2} M_{\Pl}^{2}$.  Thus,
\be
\mu \sim \frac{y}{M^2} \Lambda^{3} e^{-bT} \sim y \, m_{3/2} \frac{M_{\Pl}^{2}}{M^2}
\ee
So, for $M \sim M_{\Pl}$, $y \sim 10^{-2}$ is required to recover a $\mu$ of order the weak scale.  The low energy effective theory described by this mechanism for the $\mu$ term is simply the MSSM with mirage mediation.  

%

%

\section{Conclusions}

The mirage messenger scale depends crucially on the way that the SUSY-breaking sector couples to the modulus.  By explicitly including the light fields in the uplift sector, we have been able to explore the plausibility of different mirage scales.  If mirage mediation is to solve the little hierarchy problem, SUSY breaking must interact with the modulus field in a peculiar way such that the uplift sector becomes strongly coupled in the decompactification limit.  Only then can an effective messenger scale near the weak scale be realized.  

Mirage mediation would stand as a compelling solution
to the hierarchy problem if physics ensuring the form of the K\"ahler potential in
Section~\ref{sec:uplift} were found.  Not only would the little hierarchy problem be
ameliorated, but the $\mu/B\mu$ problem could be naturally solved, assuming
that the $\mu$ term arose from the same dynamics that generated a
superpotential for $T$.  It is interesting that a solution to the $\mu$
problem, typically a bane for theories with a large $m_{3/2}$, is
available precisely in the case where the hierarchy problem is best
solved.

Given a compelling reason for $\alpha=2$, it would then be worthwhile
to weigh cosmological issues.  In particular, \cite{NillesCCB} has noted
that a mirage spectrum at the weak scale yields tachyonic scalar masses when run
up to high energies.  Finite-temperature effects should protect against
charge/color breaking, but some might blanch at the idea of tachyonic
boundary conditions.  At zero temperature, any false vacuum is far
enough away in field space to be harmless.  It is also interesting to further
investigate the cosmology of the gravitinos and moduli in these theories.
For heavy enough masses, the gravitinos decay long-enough before big bang nucleosynthesis to
prevent a problem, but there is a direct tension between the gravitino mass and
notions of naturalness.  While it seems possible to push the
gravitinos heavy enough without completely spoiling naturalness, it would
be comforting if the cosmological difficulties could be 
completely decoupled, perhaps by incorporating the ideas of \cite{MarkusSwoS}.

\acknowledgments{We thank N. Arkani-Hamed for numerous discussions and G. Villadoro for supergravity advice.  This work is supported by the
DOE under contract DE-FG02-91ER40654.}

\appendix

\section{Series Expansion for $\alpha$}
\label{app:alpha}

In this appendix, we calculate the ratio of $F_\phi$ to $M_0 = F_T/(T+T^\dagger)$ for the supergravity Lagrangian in \eq{eq:sugralang}, using 
\be
\Omega = \Omega(X, X^\dagger,T+T^\dagger), \qquad W = a(X) e^{-bT} +  c(X).
\ee
We treat $\Omega$, $a$, and $c$ as general functions.  The constant $b$ is real and we use units where $T$ has dimensions of length and $X$ has dimensions of mass.  

As argued in the text, we are interested in expanding $F_\phi/ M_0$ in a power series in $\epsilon = 1/\langle b \Rej{T} \rangle$, and we are particularly interested in $\alpha$ defined as
\be
\frac{F_\phi}{M_0} = \frac{\alpha}{\epsilon} + \mathcal{O}(\epsilon^0).
\ee 
Recall that when $1/\epsilon \sim \log M_\Pl/ m_{3/2}$, then $\alpha = 2$ is the crucial value to achieve low fine-tuning.  To calculate the value of $\alpha$, we solve the equations of motion for $F_\phi$, $F_T$, $F_X$, and $T$ as series in $\epsilon$, and tune the cosmological constant to zero order by order in perturbation theory.  Note that the $X$ equation of motion is irrelevant for calculating $\alpha$, though in specific cases of interest, we will use the $X$ equation of motion to check that $X$ is properly stabilized.

The scalar potential is ($i= T, X$)
\be
V = F_\phi^\dagger F_\phi \Omega + \left( F_\phi^ \dagger F_i \Omega_i + \mathrm{h.c.} \right) + F_i^\dagger F_j \Omega_{i^\dagger j} + \left( 3 F_\phi W + F_i W_j + \mathrm{h.c} \right),
\ee
where subscripts on functions indicate partial derivatives.  Instead of solving for $T$ directly, it is more convenient to solve for $e^{-bT}$ as a power series in $\epsilon$, so note that
\be
\frac{\partial}{\partial T} e^{-bT} = \frac{-2 \epsilon}{T+ T^\dagger}e^{-bT}, \qquad \frac{\partial^2}{\partial T^2} e^{-bT} = \left(\frac{2 \epsilon}{T+ T^\dagger} \right)^2 e^{-bT}.
\ee
Solving the equations of motion to leading order, we find
\bea
F_\phi & =  & \frac{3 c^* \left( h^* \Omega_X - \Omega_{X^\dagger X} \right)}{\Omega \, \Omega_{X^\dagger X} - |\Omega_X|^2}+ \mathcal{O}(\epsilon), \\
F_X & = &   \frac{3c^* \left(\Omega_{X^\dagger} - h^* \Omega \right)}{\Omega \, \Omega_{X^\dagger X} - |\Omega_X|^2}+ \mathcal{O}(\epsilon), \\
\frac{F_T}{T+T^\dagger} & = & \epsilon \, \left(F_\phi - F_X \frac{\Omega_{X^\dagger X T} F_X^\dagger + \Omega_{XT} F_\phi^\dagger}{2 \Omega_{X^\dagger T} F_X^\dagger + 2 \Omega_T F_\phi^\dagger} + F_X \frac{a_X}{2a} \right) + \mathcal{O}(\epsilon^2), \\
\frac{e^{-bT}}{T+T^\dagger} & = & \epsilon \, \frac{F_\phi^\dagger \Omega_T + F_X^\dagger \Omega_{X^\dagger T}}{2a}+ \mathcal{O}(\epsilon^2) \label{eq:exptoleadingorder},
\eea
where as in the text we have defined
\be
h = \frac{c_X}{3c}.
\ee
The cosmological constant is
\be
\label{eq:finetunecc}
V_0 = 3c \left( F_\phi +  h F_X  \right) + \mathcal{O}(\epsilon),
\ee
and tuning the CC to zero in perturbation theory by adjusting (for example) $h$, we find
\be
\alpha = \frac{2}{2-\beta - \gamma}, \qquad \beta = \frac{1}{h} \left( \frac{h^* \Omega_{XT} - \Omega_{X^\dagger X T}}{\Omega_{X^\dagger T}-h^*\Omega_T} \right), \qquad \gamma = \frac{1}{h}\frac{a_X}{a}.
\ee

In \eq{eq:naturalbtsize}, the value of $1/\epsilon$ was estimated by simply comparing the $ae^{-bT}$ and $c$ terms in the superpotential.  From \eq{eq:exptoleadingorder} and using the fine-tuning condition in \eq{eq:finetunecc} we see that to leading order, the $T$ equation of motion gives
\be
\label{eq:appendixbtvalue}
e^{-bT} = \frac{F_\phi^\dagger \left(h^* \Omega_T - \Omega_{XT}^\dagger \right)}{a b h^*}.
\ee
For $M_\Pl \sim (T+ T^\dagger) \Omega_T$, $\Omega_{XT}^\dagger = 0$, $a \sim M_\Pl^3$ and $F_\phi \sim m_{3/2}$, we see that ``naturally''
\be
\langle b T \rangle \sim \log \frac{M_\Pl}{m_{3/2}},
\ee
where we have used the fact that the solution to $e^w / w = f$ is approximately $w = \log f$ for large $f$.

Finally, we calculate $\alpha$ for a field $X$ with $R$-charge $2$.  In order to stabilize $X$, \eq{eq:noomegaxt} must be augmented with higher order terms to the K\"ahler potential.  Let $M_*$ be the fundamental scale and $(s M_*)$ be the scale that sets the size of higher dimension operators involving $X$, with $s \ll 1$ by assumption.  The K\"ahler potential and superpotential take the form
\be
\Omega = -d_1 M_*^2 \left(M_* (T+T^\dagger) \right)^{n_0} + \left(  d_2 X^\dagger X - d_3 \frac{(X^\dagger X)^2}{s^2 M_*^2}  \right)  \left( M_*(T+T^\dagger) \right)^{r_X n_0} + \cdots,
\ee
\be
W = a e^{-bT} + c + 3 c h X,
\ee
where $d_i$ are dimensionless real parameters expected to be $\mathcal{O}(1)$, and we are already anticipating that $\langle X \rangle \simeq 0$.  Solving the $X$ equations of motion to leading order in $\epsilon$ and $s^2$ and tuning the CC to zero
\be
\langle X \rangle = s^2 \, \frac{d_2 h^* M_*^2}{d_3}  + \mathcal{O}(s^4),
\ee
and to leading order in $\epsilon$ and second order in $s^2$
\be
\beta = -r_X + h \langle X \rangle \frac{r_X (r_X-1)}{2} , \qquad \gamma = 0.
\ee
For sufficiently small values of $s$, order $\epsilon$ corrections to $F_\phi / M_0$ are as important as these order $s^2$ corrections, so we have seen that to the desired accuracy we can achieve
\be
\alpha = \frac{2}{2+r_X}.
\ee
As long as $d_i > 0$ and $r_X < 1$, then it is straightforward to check that all fields are stabilized and the theory contains no ghosts or tachyons.

\end{document}